\documentclass[twocolumn,apjl]{emulateapj}
\usepackage{color}

\def\mbf#1{\mbox{\boldmath ${#1}$}}
\def\Alfven{Alfv\'{e}n~}

\begin{document}

\title{Atmospheric escape by magnetically driven wind from gaseous planets}

\author{Yuki A. Tanaka$^{1}$, Takeru K. Suzuki$^{1}$ 
\& Shu-ichiro Inutsuka$^{1}$}
\email{tanaka.yuki@b.mbox.nagoya-u.ac.jp}
\altaffiltext{1}{Department of Physics, Nagoya University,
Nagoya, Aichi 464-8602, Japan}

\begin{abstract}
We calculate the mass loss driven by MHD waves from hot Jupiters by using MHD simulations in one-dimensional flux tubes. 
If a gaseous planet has magnetic field, MHD waves are excited by turbulence at the surface, dissipate at the upper atmosphere, and drive gas outflows.
Our calculation shows that mass loss rates are comparable to the observed mass loss rates of hot Jupiters, therefore it is suggested that gas flow driven by MHD waves can plays an important role in the mass loss from gaseous planets.
The mass loss rate varies dramatically with radius and mass of a planet: a gaseous planet with a small mass but with inflated radius produces very large mass loss rate.
We also derive an analytical expression for the dependence of mass loss rate on planet radius and mass that is in good agreement with the numerical calculation.
The mass loss rate also depends on the amplitude of velocity dispersion at the surface of a planet.
Thus we expect to infer the condition of the surface and the internal structure of a gaseous planet from future observations of mass loss rate from various exoplanets.
\end{abstract}

\section{Introduction}
A lot of exoplanets have been found, and many of them are considered to be gaseous giant planets like Jupiter.
In particular, some of them orbit very close ($< 0.1$\,AU) to the central stars and they are commonly known as hot Jupiters.
Transit observations of exoplanets give us not only the information of the radius of  a planet and the orbital period, but also the information of atmospheric structure and composition by spectroscopic observations.
For example, the absorption in the H Ly$\alpha$ line and the NaI D line is detected at HD 209458b \citep{cha02,vid03}.
Detection of H Ly$\alpha$ is also confirmed in other planets, such as HD 198733b \citep{lec10}, 55 Cnc b \citep{ehr12}.
The radius of HD 209458b measured in the NaI D line is not so different from the apparent radius at other wavelengths, but the radius measured in the H Ly$\alpha$ line is several times larger than the radius of the planet \citep{vid03}.
The surface temperature of HD 209458b is expected to be approximately 750\,K, but the scale height at this temperature is too small to explain the extended upper atmosphere.
Therefore, the observation suggests that the temperature of the extended upper atmosphere is much higher than that of the surface \citep{yel04}.
Recently transit observations by an X-ray space telescope show that the apparent radius of the hot Jupiter in the X-ray is few times larger than the radius of the planet \citep{pop13}.
In addition transmission spectra during primary transit and reflectance spectra around secondary eclipse probe the composition and temperature-pressure structures of the atmosphere of exoplanets \citep{mas09,mas10}.

Furthermore, transit observations strongly suggest the existence of mass loss from hot Jupiters.
The first detection of mass loss from a hot Jupiter is the transit observations of HD 209458b in UV band \citep{vid03}.
The high temperature atmosphere of hydrogen atoms that is escaping from the planet is supposed to make a cometary tail-like structure, hence dimming of a central star by escaping atmosphere is observed in UV band, especially in the Ly$\alpha$ line.
The mass loss rate is estimated by the absorption of the Ly$\alpha$ line and its value reaches at least $10^{10}\,{\rm g\,s^{-1}}$ \citep{vid03}.
Since this value is model dependent and not directly observed, an actual value of the mass loss rate is still uncertain.
It is suggested that the mass loss rate can be either larger by several orders of magnitude \citep{vl04}, or lower, for example gas trapping by closed magnetic field of the planet\citep[e.g.,][]{yel04}.

The observations of transit by H Ly$\alpha$ showed that the maximum velocity of the escaping gas is as fast as $\sim 100\,{\rm km\,s^{-1}}$ \citep{vid03}.
It is important to understand the mass loss from a planet because it will strongly affect the evolution of a planet.
However, the detailed mechanism of the strong mass loss from exoplanets is still unknown.
Various models on the structure of the upper atmosphere and the mechanism of mass loss have been developed.
For example,
energy limited escape by X-ray and extreme ultraviolet irradiation from a central star are proposed \citep{lam03}. 
This is the mechanism that certain amounts of energy of X-ray and extreme ultraviolet irradiation from the central star heat the upper atmosphere through photodissociation and photoionization, and
drives mass loss from the upper atmosphere.
A model that includes X-ray and extreme ultraviolet heating with photochemistry showed the temperature of the upper atmosphere of the irradiated gaseous planets become $\gtrsim10000\,(\rm K)$, and obtained mass loss rate is $\sim10^{10}\,{\rm g\,s^{-1}}$ \citep{yel04,yel06}.
Also, several models demonstrate that the velocity of the escaping atmosphere can be supersonic, and hydrodynamic escape dominates over Jeans escape \citep[e.g.,][]{tia05,gar07,mur09}.
Most of these models that include the heating of X-ray and extreme ultraviolet suggest the mass loss rate from hot Jupiters become $10^{9} - 10^{11}\,{\rm g\,s^{-1}}$.
X-ray and extreme ultraviolet radiation is strong in the early evolutionary phase of the central star, therefore hot Jupiters may lose the most of their masses by the XUV-driven atmospheric escape in the early phase of a system.
It is possible for a hot Jupiter to lose its entire envelope and remain only with the solid core, so the atmospheric escape from hot Jupiters may affect the population of the close-in planets \citep{kn14}.
Additionally, several previous works suggest that an effect of radiation pressure of the central star on the escaping atmosphere, and an effect of charge exchange between the escaping atmosphere and the stellar wind are important to explain the observed spectrum features by using three dimensional particle simulation \citep{hol08,eke10,bal13}.

Magnetic fields of gaseous planets are thought to be also important for the mass loss rate and the structure of the atmosphere.
Recently, several authors studied the effects of the planetary magnetic fields.
For example, magnetically controlled outflows that are launched from polar region of the planets \citep{ada11}, and numerical calculations on upper atmospheres of hot Jupiters with the magnetic fields \citep{tra11,tra14}. They also discuss the effects on the transit depth and the loss rate of the angular momentum by the planetary magnetic fields \citep{tra14}.
The effects of Ohmic dissipation in the atmosphere and internal structure of hot Jupiters are also studied\citep{bs10,per10,bat11,hc12,men12,wl13,rs14}.

However, effects of planetary magnetic field and disturbance at the surface of a planet on a mass loss rate are not investigated previously.
Stellar winds from intermediate- and low-mass stars like solar wind, are typical examples of a mechanism of mass loss due to the magnetic field.
The origin of the energy that accelerates the solar wind is supposed to be the energy of the magneto-convection and turbulence at the surface.
Mass loss by the same driving mechanism at the sun can occur in exoplanets because they are expected to have their own magnetic field and strong convection.
We calculate the mass loss driven by magnetic field from gas giants, especially from hot Jupiters by 1D MHD simulations, and analyze the dependence of the mass loss rate and the atmospheric structure on various properties of gaseous planets.
As a result, the amount of magnetically driven wind can be very large, and it can play an important role in the mass loss from gaseous planets.

In section 2 we represent our calculation method.
In section 3 we describe the calculation result, especially the dependence of the mass loss rate on the velocity dispersion at the surface of a planet, radius , and mass of a planet.
In section 4 we give summary and discussion, where we derive an analytical expression of the dependence between the mass loss rate and parameters, and compare it with numerical simulation.
We also discuss the consistency between numerical results and observations.

\section{Numerical method}
In this paper we extend our numerical simulation code for the solar wind \citep{si05,si06} to planetary winds.
The simulation code is generally applicable to stars with a surface convective layer.
So far, we have applied it to red giant winds \citep{suz07} and young active solar-type stars \citep{suz13}. 
Hot Jupiters generally possess a surface convective layer.
Therefore, they are a candidate that our simulation code is directly applicable to. 
Before describing the detailed modeling for the planetary winds, we briefly introduce general properties of stellar winds from stars with surface convection and our simulation code. 

In stars with a surface convective layer, magnetic fields are generated by dynamo action \citep[e.g.,][]{cho95,bru04,hot12}, and various types of magnetic waves are excited \citep{ms12,ms13}. 
Among them, Alfv\'{e}n waves are an promising source which transfers the energy in the surface convection to the upper atmosphere;  Alfv\'{e}n waves, which propagate upwardly from the surface, heat up the upper corona and drive the stellar wind by various dissipation processes \citep{gol78,hp83,ter86,ks99,mat99}.

We time-dependently solve the propagation and dissipation of such MHD waves and consequent heating of the gas in a single open flux tube.
In order to take into account closed loops which cover a sizable fraction of the surface, we consider super-radially open magnetic flux tubes of which the radial magnetic field strength, $B_r$, is determined by the conservation of magnetic flux, 
\begin{equation}
B_r r^2 f(r) = B_{r,0}r_{0}^2 f_0, 
\end{equation} 
where $f(r)$ indicates an areal filling factor of open flux tubes at radial distance $r$, and a subscript '0' represents the surface. 
We use the same functional form of $f(r)$ as in \citet{suz13},
\begin{equation}
f(r) = \frac{e^{\frac{r-r_0-h_{\rm l}}{h_{\rm l}}} + f_0 - (1-f_0)/e}
{e^{\frac{r-r_0-h_{\rm l}}{h_{\rm l}}}+1}, 
\end{equation}
where $h_{\rm l}$ denotes a typical height of closed loops.
This is essentially the same form as a super-radial expansion factor which was introduced by 
\citet{kh76}.

In the 1D open flux tube, we solve the following MHD equations with radiative cooling and thermal conduction, 
\begin{equation}
\label{eq:mass}
\frac{d\rho}{dt} + \frac{\rho}{r^2 f}\frac{\partial}{\partial r}
(r^2 f v_r ) = 0 , 
\end{equation}
\begin{equation}
\label{eq:mom}
\rho \frac{d v_r}{dt} = -\frac{\partial p}{\partial r}  
- \frac{1}{8\pi r^2 f}\frac{\partial}{\partial r}  (r^2 f B_{\perp}^2)
+ \frac{\rho v_{\perp}^2}{2r^2 f}\frac{\partial }{\partial r} (r^2 f)
-\rho \frac{G M_{\star}}{r^2}  , 
\end{equation}
\begin{equation}
\label{eq:moc1}
\rho \frac{d}{dt}(r\sqrt{f} v_{\perp} - \frac{G M_{\star}}{r}) 
= \frac{B_r}{4 \pi} \frac{\partial} {\partial r} (r \sqrt{f} B_{\perp}).
\end{equation}
$$
\rho \frac{d}{dt}(e + \frac{v^2}{2} + \frac{B^2}{8\pi\rho}) 
+ \frac{1}{r^2 f} 
\frac{\partial}{\partial r}[r^2 f \{ (p + \frac{B^2}{8\pi}) v_r  
- \frac{B_r}{4\pi} (\mbf{B \cdot v})\}]
$$
\begin{equation}
+ \frac{1}{r^2 f}\frac{\partial}{\partial r}(r^2 f F_{\rm c}) 
+ q_{\rm R} = 0,
\label{eq:eng}
\end{equation}
\begin{equation}
\label{eq:moc}
\frac{\partial B_{\perp}}{\partial t} = \frac{1}{r \sqrt{f}}
\frac{\partial}{\partial r} [r \sqrt{f} (v_{\perp} B_r - v_r B_{\perp})], 
\end{equation}
where $\rho$, $\mbf{v}$, $p$, $e$, $\mbf{B}$ are density, velocity, pressure, specific energy, and magnetic field strength, respectively, and subscript $r$ and $\perp$ denote radial and perpendicular components. 
$\frac{d}{dt}$ and $\frac{\partial}{\partial t}$ denote Lagrangian and Eulerian derivatives, respectively.  
$G$ and $M_{\star}$ are the gravitational constant and the mass of a central object.
$F_{\rm c}$ is thermal conductive flux and $q_{\rm R}$ is radiative cooling, which is explained later. 
Note that the curvature effects appear as $r\sqrt{f}$ terms, instead of $r$ for the usual spherical coordinates.
We adopt 2nd-order MHD-Godunov-MOCCT scheme to update the physical quantities \citep{san99}.
Also, this calculation method for stellar winds from solar-type stars is developed to 2D calculation, and more detailed aspects of stellar winds have been investigated \citep{ms12}.

In the simulations for the solar and stellar winds \citep{si05,si06,suz07,suz13}, we set the inner boundary at the photosphere. 
For the planetary winds in this paper, we set the inner boundary at the position that gives $p_0=10^5$ dyn cm$^{-2}$ ($=0.1$ bar). 
We fix the temperature at the inner boundary to the given surface temperature, $T_0$. 
The density at the inner boundary is accordingly determined to give $p_0$. 

Since the strength and the configuration of magnetic field in hot Jupiters have large uncertainties, we set up an open magnetic flux tube referring to observation of the Sun.
Recent observations by the HINODE satellite shows that the footpoints of open flux tubes in polar regions are anchored to so-called kG patches \citep{tsu08,ito10,shi12}. 
The field strength is approximately the equipartition to the gas internal energy.
The magnetic field lines are super-radially open with an elevating altitude and the cross section is typically expands with a factor of 1000, which indicates that the filling factor of open flux tubes at the photosphere is an order of $1/1000$.
As a result, the typical field strengths in the coronal region are an order of 1G.
Applying these obtained magnetic properties on the Sun to planetary winds, we impose the radial magnetic field with the strength being the equipartition value to the gas pressure, or in other words, plasma $\beta$
\begin{equation}
\beta = \frac{8\pi p}{B^2},
\end{equation}
equals to unity at the inner boundary \citep{cs11}.
In our setups, the value that satisfies this condition is $B_{r,0}=1.59$ kG.
The filling factor at the inner boundary is set to be $f_0=1/1600$, which indicates that the average field strength contributed from open flux tube regions is $\approx 1$ G. 
The filling factor is, in principle, determined by the force balance between the outflowing wind and the magnetic field; larger wind mass flux opens up closed magnetic structure to lead to larger $f_0$. 
Therefore, we should carefully examine the obtained wind profile in comparison to the field strength in the outer region. 
In typical cases, the Alfv\'{e}n point, the location at which the Alfv\'{e}n velocity equals to the radial flow velocity, is 10-20 planetary radii. 
This indicates that the wind kinetic energy is comparable to the energy of the radial magnetic field at this location. 
The obtained Alfv\'{e}n point in units of the object's radius is quite similar to that for the solar wind, and the adopted $f_0$ is supposed to be 
reasonable.

As the standard case of our simulations, we take a hot Jupiter with the surface temperature, $T_0=1000$ K.
We inject velocity perturbations with amplitude, $\delta v_0/c_{\rm s,0}=0.2$, at the inner boundary, where $c_{\rm s}$ is sound speed.
Here, we assume a broad band spectrum of $\delta v_0$ in proportion to $1/\nu$, where $\nu$ is frequency.
For the standard case, we adopt the loop height, $h_{\rm l}=0.5 r_0$, which controls the location where the open flux tube most rapidly opens. 
When changing the surface temperature, we change $h_{\rm l}$ in proportion to $T_0$, because we expect that $h_{\rm l}$ is scaled by the pressure scale height, which $\propto T_0$.

We perform simulations of hot Jupiters with the solar metallicity and take the radiative cooling (Equation \ref{eq:eng}) from the solar abundance gas.
In the simulations for the solar and stellar winds \citep{si05,si06,suz07,suz13}, we have taken the optically thin radiative cooling for the coronal plasma \citep{lm90,sd93} and empirical radiative cooling for the chromospheric gas that takes into account optically thick effects based on observations of the solar chromosphere \citep{aa89}. 
For the simulations of hot Jupiters, we need to prescribe the radiative cooling for gas with lower temperature down to $\sim 1000$ K. 
In this paper, we adopt a simple treatment by extending the empirical cooling rate \citep{aa89}, which is proportional to density, $4.5\times 10^9\rho$ (erg cm$^{-3}$s$^{-1}$). 
We switch off the cooling when temperature becomes lower than the surface temperature, $T_0$. 
This treatment is probably too much simplified; we plan to elaborate the treatment of the cooling in our future works
(see Section 4.3).

As shown in Equations (\ref{eq:mass}) -- (\ref{eq:moc}), our treatment is based on one fluid MHD, which requires well coupling between gas and magnetic fields.
To fulfill this condition, sufficient electrons are necessary to couple field lines and weakly ionized media, although the required ionization degree can be as small as $10^{-10}$ -- $10^{-5}$ whereas the actual value is up to density.

In our calculation we assume the ideal MHD approximation.
However, it is not clear that whether the ideal MHD approximation is applicable for the calculation of hot Jupiters or not, because the temperature is not so high $\sim1000\,{\rm K}$ and ionization degree supposed to be small.
Here we show the applicability of the ideal MHD approximation for our calculations of the atmosphere of hot Jupiters following an estimate for the atmosphere of brown dwarfs by Sorahana et al.(2014).
An induction equation is expressed as follow,
\begin{eqnarray}
\frac{\partial \mbf{B}}{\partial t}=\nabla\times\left(\mbf{v}\times\mbf{B}\right)-\eta\left(\nabla\times\mbf{B}\right),
\end{eqnarray}
where $\eta$ is resistivity.
This equation describes the evolution of magnetic field, and $\eta=0$ corresponds to the ideal MHD approximation.
If the second term of the right hand side dominates over the first term, the magnetic field diffuse out and the ideal MHD approximation is no longer valid.
The most dominant origin of the resistivity in the atmosphere of hot Jupiters is the collision between electrons and neutral particles.
This resistivity depends on temperature and an ionization degree, and it can be expressed as
\begin{eqnarray}
\eta\approx200\frac{\sqrt{T\, ({\rm K})}}{x_{e}}\,({\rm cm^{2}\, s^{-1}})\label{resistivity}
\end{eqnarray}
\citep[e.g.,][]{bb94}.
To estimate the applicability of the ideal MHD approximation, it is useful to introduce a magnetic Raynolds number,
\begin{eqnarray}
R_{m}=vL/\eta .
\end{eqnarray}
$L$ is a typical length of a system.
If the magnetic Raynolds number $R_{m}$ is quite larger than unity, the ideal MHD approximation is applicable.
For the typical length of the system, we use the typical wavelength of \Alfven wave that we are injecting as a perturbation,
\begin{eqnarray}
L\sim v_{\rm A}\tau\sim c_{s}\tau.\label{typicallength}
\end{eqnarray}
$v_{\rm A}$ is the \Alfven velocity and almost same as $c_{s}$, because we assume the equipartition of gas pressure and magnetic pressure for the magnetic flux tube in our calculation.
$\tau$ is a typical timescale and given approximately by the pressure scale height of the atmosphere divided by the sound speed of the surface.
In the situation that a planet has Jupiter radius and Jupiter mass and the surface temperature is 1000 K, the typical timescale can be written as
\begin{eqnarray}
L\sim270\,{\rm km}\left(\frac{c_{s}}{2.6\,{\rm km\,s^{-1}}}\right)\left(\frac{\tau}{100\,{\rm s}}\right)\label{typl2}
\end{eqnarray}
from equation (\ref{typicallength}).
From equation (\ref{resistivity}) and (\ref{typl2}), the estimation of $R_{m}$ as follow:
\begin{eqnarray}
R_{m}=2.2\left(\frac{v}{0.5\,{\rm km\,s^{-1}}}\right)\left(\frac{\tau}{100\,{\rm s}}\right)\left(\frac{x_{e}}{10^{-8}}\right)\label{rm1}
\end{eqnarray}
where $v$ is normalized by the value of the velocity dispersion we are injecting, $\delta v=0.2c_{s}$.
According to equation (\ref{rm1}), the ideal MHD approximation is applicable to the description of the atmosphere of hot Jupiters even if the ionization degree is not so large.
The requirement condition for the ideal MHD approximation is more mild in upper atmosphere, because the amplitude of \Alfven waves become larger due to decreasing of the density of the atmosphere.
$\tau$, $L$ and many other terms can vary with the difference of the conditions, because there are a variety of properties of hot Jupiters.
For example, the scale height is significantly large in the gaseous planet with smaller mass, larger radius and hence, lower surface gravity.
In this case, typical timescale $\tau$ is larger because of large scale height, therefore the ideal MHD approximation also is better.

\section{Results}
In this section we describe parameter dependence of the mass loss rate from gaseous planets and the atmospheric structure.

\subsection{Dependence on velocity dispersion}
First, we show the relation between velocity dispersion of the magnetic field at the surface of planets and the mass loss rate from gaseous planets, and the atmospheric structure.
Perturbations at the surface excite MHD waves, and they propagate upward.
MHD waves dissipate at the upper atmosphere, then gas is heated and given momentum.
Here we assume that Poynting flux of MHD wave driven by disturbance at the surface drives gas flow from gaseous planets.
In this calculation disturbance of the magnetic field is given at the photosphere of gaseous planets.
Total energy transported by MHD wave varies with given strength of disturbance, therefore the mass loss rate should depend on the velocity dispersion.
The origin of disturbance is turbulence that is caused by the convection.
For example, the velocity dispersion of turbulence at the photosphere of the sun is about $20-30\%$ of sound speed \citep{mk11}.
The strength of turbulence at the surface of exoplanets is unknown, so we treat the value of velocity dispersion as a parameter.
We adopt its value comparable to the value at the sun, or less than it.
In the case of young gaseous planets, velocity dispersion of turbulence might be very large because cooling due to convective heat transport is expected to be active.
Therefore, the values of parameters we adopted for this calculation and the resultant mass loss rate might be underestimated.

\begin{figure}[h]
\includegraphics[]{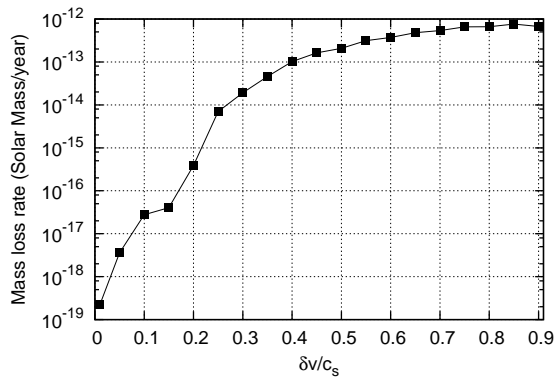}
\caption{A relation between the strength of disturbance given to the magnetic field line at the surface of planets and the mass loss rate from gaseous planets.
Horizontal axis is the velocity dispersion normalized by the sound speed at the surface of planets.
Vertical axis is the mass loss rate in units of $M_{\odot}/yr$.
}
\label{dv-massloss}
\end{figure}

A relation between the velocity dispersion at the surface of gaseous planets and the mass loss rate is shown in Fig \ref{dv-massloss}.
Mass loss rate increases with an increase of the value of the velocity dispersion, because of the larger energy deposition in the magnetic flux tube.
In the small velocity dispersion region the mass loss rate increases very rapidly with $\delta v$, but 
in the high velocity dispersion region the increase of the mass loss rate become small.
In spite of the increase of energy deposited to the magnetic flux tube the mass loss rate saturates.
This result suggests that there is an upper limit of the mass loss rate that is driven by magnetic energy.
The increase of the mass loss rate is mainly by the increase of the density in the wind, which also enhances the radiative cooling because it is in proportional to $\rho^2$ in the optically thin limit. 
As a result, a larger fraction of the input Poynting flux is lost by the radiation rather than transferred to the kinetic energy of the wind \citep{suz13}.
This is the main reason why the mass loss rate saturates for the large $\delta v$ limit. 

\begin{figure}[h]
\includegraphics[]{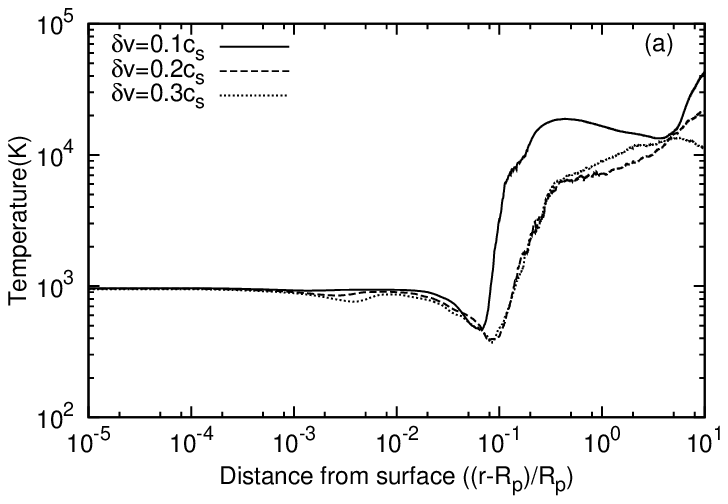}
\includegraphics[]{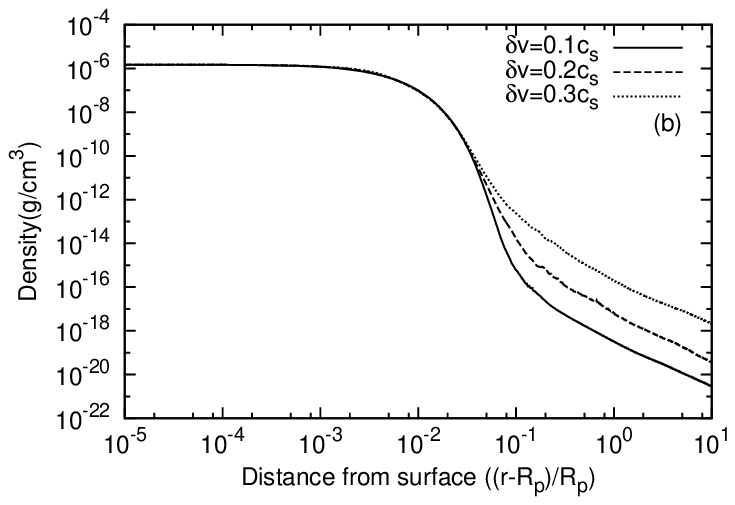}
\includegraphics[]{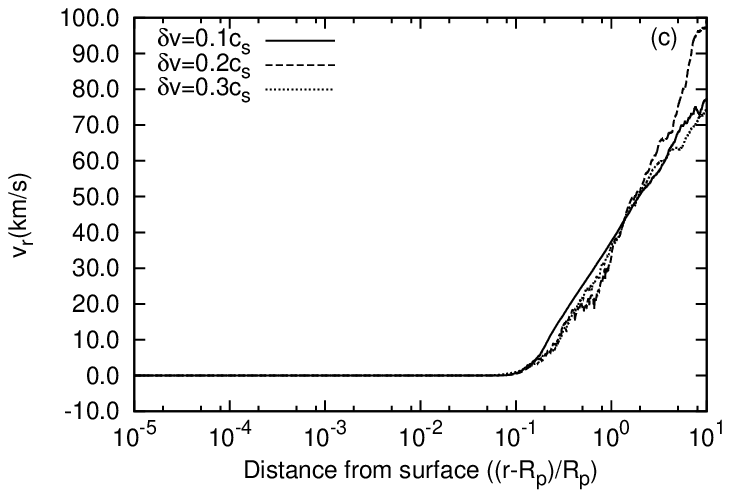}
\caption{The velocity dispersion dependence of the atmospheric structure.
(a) temperature structure, (b) density profile, and (c) radial velocity profile.
Horizontal axis denotes the distance from the surface that is normalized by the radius of the planet in logarithmic scale.
Solid, dashed, dotted lines correspond to the values of velocity dispersion, 0.1$c_{s}$, 0.2$c_{s}$, and 0.3$c_{s}$, respectively.
The atmospheric structure changes dramatically at the upper atmosphere by the dissipation of magnetic energy.
}
\label{dv-structure}
\end{figure}

The structures of temperature, density, and radial velocity of the atmosphere are shown in Fig \ref{dv-structure}.
Temperature at near surface region is approximately constant and its value is same as the surface temperature $1000\,$K, while gas is heated and the temperature increases rapidly to over $10000\,$K at the upper atmosphere.
Corona-like regions appear at the upper atmosphere of the gaseous planets, whereas the temperature here is much lower than the temperature of the solar corona ($\sim\,10^{6}\,{\rm K}$).
The density profiles of the lower atmosphere are almost same regardless of the given values of the velocity dispersion, but they changes in the regions where the temperature rises rapidly due to the dissipation of MHD waves.
If the value of the velocity dispersion is larger, larger amount of gas is uplifted and the density decreases more gradually than that of the case with the smaller velocity dispersion.

The value of the velocity dispersion and the strength of convection at the surface of exoplanets are unknown.
The origin of the velocity dispersion at the surface of the Sun is considered to be turbulence in the surface convective layer.
While in gaseous planets the similar mechanism is supposed to operate, in hot Jupiters the radiative-convective boundary is located in at a deep location ($\ge 1\,{\rm kbar}$) \citep{bur03,for07}.
However convective over-shoot and propagating waves from convection region may provide the injected $\delta v$ in our paper, whereas $\delta v/c_{s}$ might be smaller than the typical solar value because of the deeper radiative-convective boundary.
On the other hand, we cannot imagine that there is no flow on the surface of rotating gaseous object irradiated from one side.
Indeed, recent numerical studies for atmospheric circulation on hot Jupiters suggest that equatorial wind speed can reach $2-5\,{\rm km\,s^{-1}}$ \citep{sg02,cs05,dl08}.
These results imply that supersonic atmospheric flow can exist on the surface of hot Jupiters, therefore turbulence may be created in the atmosphere of hot Jupiters.
In the following section, we adopt $\delta v=0.2c_{s}$ as a typical value of the velocity dispersion at the surface of  gaseous planets.


\subsection{Dependence on planet radius}
Next we describe the relation between the radius of gaseous planets and the mass loss rate.
The observed value of the radius of a hot Jupiter varies from 0.8$R_{J}$ to 2$R_{J}$.
We calculate the dependence of the mass loss rate and the atmospheric structure on the radius of the planet.
Here we adopt $1000\,$K for the surface temperature, Jupiter's mass for the mass of the planet.

\begin{figure}[h]
\includegraphics[]{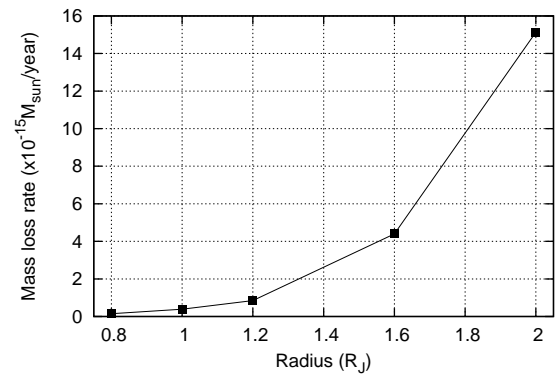}
\caption{
Relations between the radius of planet and the mass loss rate.
Horizontal axis is the radius of the planet normalized by the radius of Jupiter.
The mass loss rate increases rapidly with an increase in the radius.
}
\label{radius-massloss}
\end{figure}

Change of the mass loss rate with the radius of a planet is shown in Fig \ref{radius-massloss}.
The mass loss rate is small when the radius of a planet is small, but it increases dramatically with an increase of the radius of a planet.
As shown in Fig. \ref{radius-massloss}, the mass loss rate increases by an order of magnitude when the radius is doubled.
Therefore, hot Jupiters with inflated radii are expected to have larger amount of the mass loss rate compared with that of ordinary hot Jupiters, if the values of the surface temperature, mass of planets, and velocity dispersion at the surface are not very different.

\begin{figure}[h]
\includegraphics[]{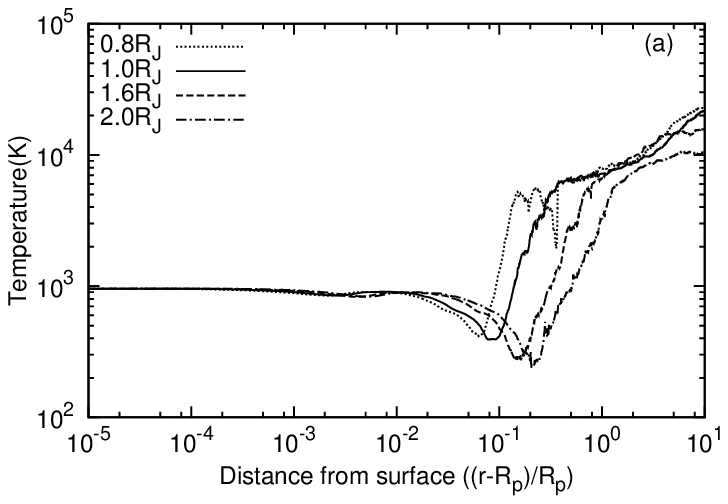}
\includegraphics[]{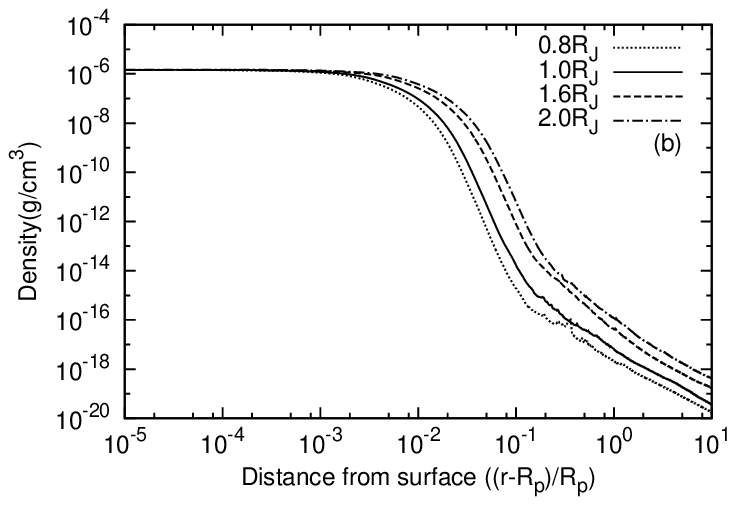}
\includegraphics[]{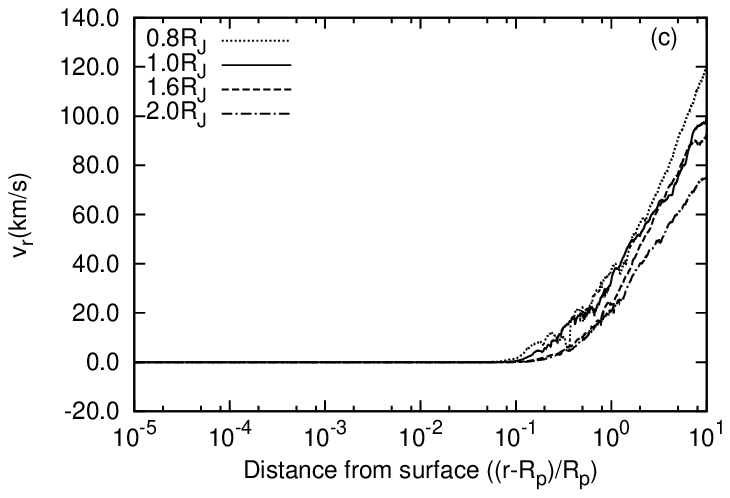}
\caption{The planet radius dependence of the atmospheric structure.
(a) temperature structure, (b) density profile, and (c) radial velocity profile.
Horizontal axis and vertical axis are same as Fig \ref{dv-structure}.
Dotted, solid, dashed, dotted-dashed lines correspond to radius of planets, 0.8$R_{J}$, 1$R_{J}$, 1.6$R_{J}$, and 2.0$R_{J}$ respectively.}
\label{radius-structure}
\end{figure}

Fig \ref{radius-structure} shows the relations between the atmospheric structures and the planet radius.
The temperatures of the upper atmospheres are heated up to $\sim 10^{4}\,{\rm K}$ in all cases, but heating starts at lower altitude in the case that radius is smaller.
The speeds of the planetary winds are not so different in all cases, but it is slightly faster in smaller radius planet because of the difference of the escape velocities that is determined by surface gravity.
We give a detailed description of the dependence between the radius of a planet and the mass loss rate in section \ref{paradep}.

\subsection{Dependence on planet mass}
Here we describe the relation between the mass of a planet and the mass loss rate.
To date many exoplanets with various mass have been detected from rocky planets in a range of the Earth-size mass to gas giant planets with super-Jupiter mass.
Here we change the mass of planets from 0.3$M_{J}$ to 1.5$M_{J}$ and calculate the mass loss rate and the atmospheric structure.
The surface temperature is set to $1000\,$K, and the radii of planets are set to Jupiter's radius.

\begin{figure}[h]
\includegraphics[]{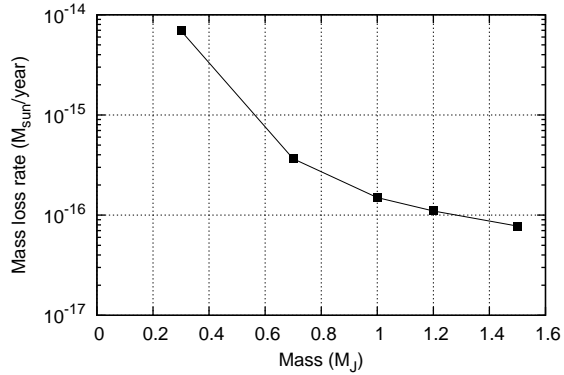}
\caption{A relation between the mass of planets and the mass loss rate.
Horizontal axis is the mass of planets normalized by the mass of Jupiter.
The mass loss rate decreases with an increase in mass.
}
\label{mass-massloss}
\end{figure}

\begin{figure}[h]
\includegraphics[]{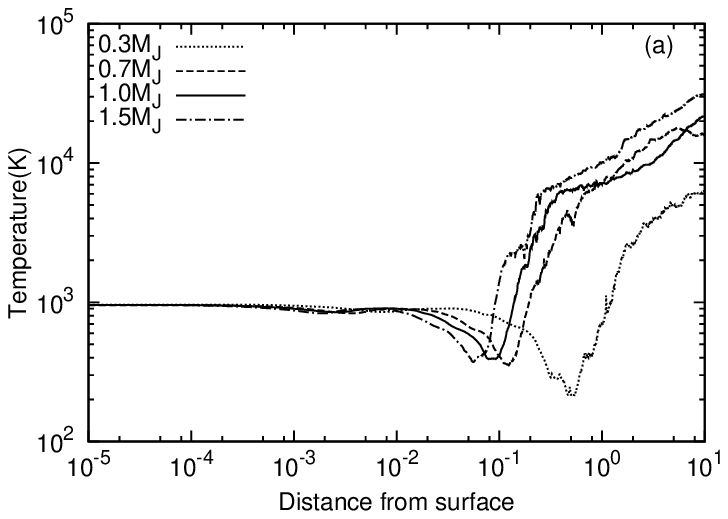}
\includegraphics[]{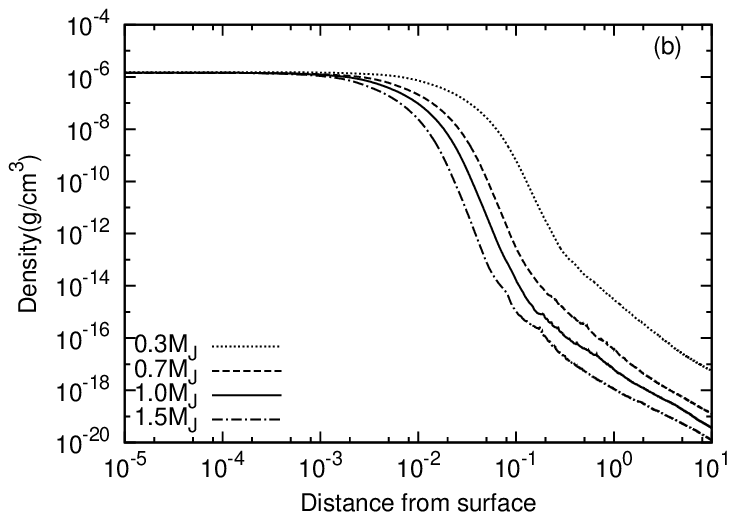}
\includegraphics[]{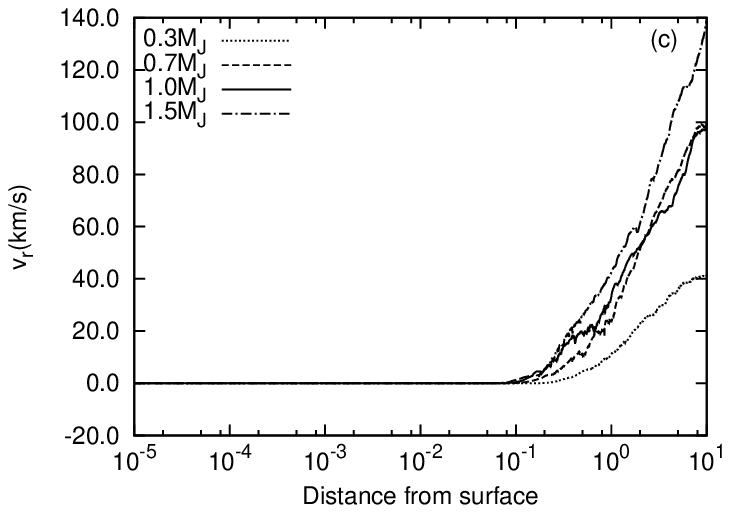}
\caption{
The mass dependence of the atmospheric structure.
(a) temperature structure, (b) density profile, and (c) radial velocity profile.
Horizontal axis and vertical axis are the same as Fig \ref{dv-structure}.
Dotted, dashed, solid, dot-dashed lines correspond to the cases with $M_{p}=0.3M_{J}$, $0.7M_{J}$, $1.0M_{J}$, and $1.5M_{J}$, respectively.
}
\label{mass-structure}
\end{figure}

The relation between the mass of planets and the mass loss rate is shown in Fig \ref{mass-massloss}.
As the mass of planets increases, gas flow from planets become difficult to blow out, which is understandable.
The structures of temperature, density profile, and radial velocity profile are shown in Fig \ref{mass-structure}.

As shown in the bottom panel of Fig \ref{mass-structure}., the acceleration of planetary winds depend on the mass of planets.
The speed of the planetary wind from $0.3M_{J}$ planet is particularly small because the wind velocity is roughly scaled by the escape velocity $\propto \sqrt{M_{p}/R_{p}}$.
However, the mass loss rate from the planet is very large in spite of its slow wind, because the density of the planetary wind is very large as shown in the middle panel of Fig \ref{mass-structure}.
Slow and dense planetary wind blows out from a lighter planet, and fast and low density wind blows out from a heavier planet.

\subsection{Dependence on radius and mass}
\begin{figure}[h]
\includegraphics[]{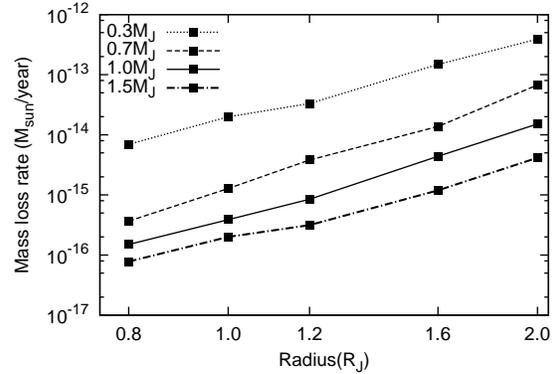}
\caption{
The relation between the radius of a planet and the mass of a planet, and the mass loss rate.
Horizontal axis is the radius of planets normalized by Jupiter's radius, and vertical axis is the mass loss rate.
Note that both axes are logarithmic scale.
Dotted, dashed, solid, dotted-dashed lines correspond to the mass, $0.3M_{J}$, $0.7M_{J}$, $1.0M_{J}$, $1.5M_{J}$, respectively.
}
\label{radiusmass-massloss}
\end{figure}

We described the result of the calculation that changed only one parameter previously.
Here we show the dependence of the mass loss rate on both the radius of a planet and the mass of a planet.
From previous discussion, the mass loss rate increases with an increase in the radius of a planet, and increases with a decrease in the mass of a planet.
Both axes in Fig \ref{radiusmass-massloss} are logarithmic scale.
Each line is corresponding to the mass of planets, and they are approximately parallel.

\section{Summary $\&$ Discussion}

\subsection{Parameter dependence of mass loss rate}\label{paradep}
\begin{figure}[]
\begin{center}
\includegraphics[]{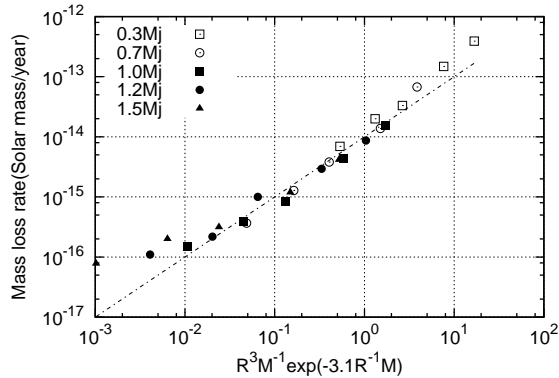}
\end{center}
\caption{
Comparison between analytical expression and numerical simulation of mass loss rate.
Horizontal axis is $R^{3}M^{-1}\exp\!\left(-3.1R^{-1}M\right)$ , factor $-3.1$ is suggested by fitting, and vertical axis is the mass loss rate.
The dot-dashed line correspond to $\dot{M}=\frac{R^{3}}{M}\exp\!\left(-3.1\frac{M}{R}\right)$
}
\label{allfit}
\end{figure}

Here we discuss the dependence of the mass loss rate from gaseous planets on the radius and mass of planets.
We assume that the surface temperatures of planets are $1000\,$K, and the value of the velocity dispersion at the surface is constant, 20$\%$ of the sound speed for simplification.
As described in Section 3, the mass loss rate from gaseous planets increases with the radius, and decreases with the mass.
This can be understood by the dependence of the scale height of the atmosphere on the surface gravity as follows.
In our model, MHD waves caused by disturbance of the atmosphere at the surface of the planet propagate into upper atmosphere, and the planetary wind is driven by the dissipation of energy of MHD waves.
As shown in Fig. \ref{dv-structure}, Fig. \ref{radius-structure}, and Fig. \ref{mass-structure}, the region where the atmosphere is heated and accelerated rapidly is located at few $\%$ of planetary radius above the surface of the planet. 
The mass loss rate from the planet depends on the density of the region where the planetary wind is accelerated, and its density varies with the scale height of the atmosphere.
The mass loss rate should increase with the scale height of the atmosphere, because the density of the upper atmosphere with large scale height is large.
The mass loss rate and other parameters are expected to be related by the following equation,
\begin{eqnarray}
\label{eq:masslossprop}
\frac{1}{2}\dot{M}v_{esc}^{2}\propto4\pi R^{2}\rho\!\left(r_{c}\right) v_{w}\langle\delta v^{2}\rangle
\end{eqnarray}
where $v_{esc}$, $v_{w}$, and $\delta v$ are corresponding to escape velocity at the surface of an object, Alfven velocity, and velocity dispersion, respectively.
The right hand side is energy flux at the transonic point where wind velocity coincides with sound velocity $r_{c}$ of the planetary wind, and the left hand side is kinetic energy transported by the planetary wind per unit time.
Planetary wind is supposed to stream out from open flux regions which probably cover a fraction of the surface.
In addition, an only small fraction of the injected energy from the surface is transferred to the final kinetic energy of the wind after suffering from the reflection of wave and radiative energy loss \citep[e.g.,][]{suz13}.
By expressing these corrections as $f_{c}$, equation (\ref{eq:masslossprop}) can be written as follow,
\begin{eqnarray}
\frac{1}{2}\dot{M}v_{esc}^{2}=f_{c}\cdot4\pi R^{2}\rho\!\left(r_{c}\right) v_{w}\langle\delta v^{2}\rangle.
\end{eqnarray}
Escape velocity is 
\begin{eqnarray}
v_{esc}^{2}=\frac{2GM}{R}.
\end{eqnarray}
The value in the sun is $f_{c}\sim10^{-5}$, and the value of $f_{c}$ is expected to range $10^{-3}-10^{-6}$\citep{suz13}.
From the results of our calculations, values of $f_{c}$ in hot Jupiters vary from $\sim10^{-4} - 10^{-6}$ depending on the parameters.
A typical value is a few times $10^{-6}$, for example, $f_{c}\simeq3.21\times10^{-6}$ in the case that $R=R_{J}$, $M=M_{J}$, the surface temperature is $1000\,{\rm K}$, and the velocity dispersion at the surface is $0.2c_{s}$.
To estimate the mass loss rate from gaseous planets, we have to estimate the density profile of the atmosphere because the density at the transonic point controls the mass loss rate.
First, we assume that the atmosphere is in hydrostatic equilibrium.
Although the region where the planetary wind is accelerated by the dissipation of MHD wave energy is no longer hydrostatic equilibrium, the hydrostatic density structure is still a reasonable approximation in the sub-sonic region.
The density profile of the atmosphere in hydrostatic equilibrium determines the density in the acceleration point, and it influences the mass loss rate.
The equation of hydrostatic equilibrium is written as
\begin{eqnarray}
\label{eq:hydroeq}
\frac{1}{\rho}\frac{dp}{dr}+\frac{GM_{p}}{r^{2}}=0.
\end{eqnarray}
$p$ is a pressure of atmosphere, $\rho$ is a density, and $M_{p}$ is a mass of a planet.
By assuming isothermal flow, the density profile is derived as
\begin{eqnarray}
\frac{\rho}{\rho_{0}}&=&\exp\!\left(-\frac{GM_{p}}{c_{s}^{2}}\left(\frac{1}{R}-\frac{1}{r}\right)\right)\nonumber\\
&=&\exp\!\left(-\frac{r-R}{H_{0}}\frac{R}{r}\right)
\end{eqnarray}
by using the scale height $H_{0}=N_{A}k_{B}T/\mu g_{0}$.
$\rho_{0}$ is the density at the surface.
This is an approximate expression for the density profile of the atmosphere.
By using these equations, we can obtain an expression for the mass loss rate, 
\begin{eqnarray}
\dot{M}\propto\frac{R^{3}}{M}\exp\!\left(-\frac{G}{c_{s}^{2}}\frac{r_{c}-R}{r_{c}}\frac{M}{R}\right).\label{masslosseq}
\end{eqnarray}
The acceleration point of planetary wind $r_{c}$ is about twice as large as planetary radius, and $\left(r_{c}-R\right)/r_{c}$ is order unity.
We fit the factors with assuming $\left(G/c_{s}^{2}\right)\left(r_{c}-R\right)/r_{c}$ is constant for simplification. 
As a result, the dependence of the mass loss rate on the radius with the same mass can be expressed as the following equation,
\begin{eqnarray}
\dot{M}=1.109\times10^{-14}\left(\frac{R}{R_{J}}\right)^{3}\exp\!\left(-3.27397\frac{R_{J}}{R}\right)\,{\rm g\,s^{-1}}.
\end{eqnarray}
This analytically-derived parametric dependence of the mass loss rate is shown in Fig \ref{allfit}. 
Good agreement to our numerical simulation can be seen in the figure.

\subsection{Comparison with Observation}

Although observations of the mass loss rate from hot Jupiters are limited, a lower limit of the mass loss rate is estimated from light curve during transit.
According to the observation and analysis, it is estimated that lower limit of the mass loss rate from HD 209458b, that is considered a typical hot Jupiter, is  $10^{10}\,{\rm g\,s^{-1}}$ \citep{vid03}.
Another transit observation also suggests that the mass loss rate from HD 209458b is in the range $(8-40)\times10^{10}\,{\rm g\,s^{-1}}$ \citep{lin10}.
Note that these values of the mass loss rate are model-dependent, therefore they are not directly observed values.
In our calculation, the values of the mass loss rate are $1.8\times10^{9}\,{\rm g\,s^{-1}}$, $2.4\times10^{10}\,{\rm g\,s^{-1}}$, and $1.2\times10^{12}\,{\rm g\,s^{-1}}$ with the velocity dispersion are $\delta v=0.1c_{s}$, $0.2c_{s}$, and $0.3c_{s}$, respectively, their values are comparable with the estimation of the lower limit of the observation.
The mass loss rate in the case of $\delta v=0.2c_{s}$ is the most similar value to the
estimated value and the previous work.
Additionally, the velocity component as fast as $\sim100\,{\rm km\,s^{-1}}$ is observed \citep{vid03}.
Our calculations show that the velocity of escaping atmosphere is supersonic in all cases, and the values exceed $100\,{\rm km\,s^{-1}}$ in some cases.
These results are consistent with the observations.
The mass loss rate depends on the velocity dispersion, surface temperature, radius and mass of a planet, but these quantities can be determined by observations in near future, except for the velocity dispersion.
In other words, we can estimate amplitude of the velocity dispersion at the surface of gaseous planets if our mechanism explains the observed mass loss rate from gaseous planets.

\subsection{Future Work}

In this paper, we calculate the mass loss rate from hot Jupiters and discuss its dependence on the velocity dispersion at the surface, radius, and mass of a planet with the constant surface temperature.
The adopted surface temperature of a planet is $1000\,$K that is considered to be a typical temperature of hot Jupiters.
To calculate the mass loss rate from lower temperature planets by decreasing the surface temperature, we should refine our calculation with more detailed thermal physics.
Then we can discuss the dependence of the mass loss rate on the surface temperature in near future.
The surface temperature of a planet strongly depends on irradiation from the central star especially in a region near the star, therefore the surface temperature should be correlated with the semi-major axis of the planet.
Calculation of  the mass loss rate not only from a hot Jupiter, but also from a gaseous planet located at several AU from a central star is our next work.

The present treatment of the radiative cooling particularly for gas with $T<10^4$ K is crude. 
As described in Section 2, we adopt the empirical radiative loss function for the solar choromosphere  \citep{aa89} for the gas with $4000$ K $\lesssim T \lesssim 10^4$ K, in which the main coolants are HI, MgII, CaII, and FeII \citep{ver81} 
The cooling rates for these species are smaller for gas with lower temperature \citep{aa89b}.
On the other hand, molecules (e.g. CO, SiO, CS, OH, H$_2$O, etc.) supersedes them as dominant coolants \citep{tsu67, tsu73}. 
Although the cooling rate of these molecules seems roughly comparable to that of the main coolants for the solar chromosphere \citep[e.g.,][for SiO]{muc87}, the precise 
cooling rate depends on the actual structure of an atmosphere \citep{muc87}. 
At present, we cannot tell whether our simplified treatment for the cooling overestimates or underestimates the radiative loss; in future study, we need to incorporate these species with radiative transfer in a self-consistent manner into our dynamical simulations.

\section*{Acknowledgement}
We thank Hiroyuki Kurokawa for giving us helpful advice to improve the paper.
This work was supported in part by Grants-in-Aid for Scientific Research from the MEXT of Japan, 22864006 (TKS).


\begin{thebibliography}{99}

\bibitem[Adams (2011)]{ada11}
Adams, F. C. 2011, \apj, 730, 27

\bibitem[Anderson \& Athay (1989a)]{aa89}
Anderson, L. S. \& Athay, R. G. 1989, \apj, 336, 1089

\bibitem[Anderson \& Athay (1989b)]{aa89b}
Anderson, L. S. \& Athay, R. G. 1989, \apj, 346, 1010

\bibitem[Ayres (1981)]{ayr81}
Ayres, T. R. 1981, \apj, 244, 1064

\bibitem[Batygin \& Stevenson (2010)]{bs10}
Batygin, K. \& Stevenson, D. J. 2010, \apjl, 714, L238

\bibitem[Batygin et al.(2011)]{bat11}
Batygin, K., Stevenson, D. J., \& Bodenheimer, P. H. 2011, \apj, 738, 1

\bibitem[Blaes \& Balbus (1994)]{bb94}
Blaes, O. M. \& Balbus, S. A. 1994, \apj, 421, 163

\bibitem[Bourrier \& Lecavelier des Etangs (2013)]{bal13}
Bourrier, V. \& Lecavelier des Etangs, A. 2013, \aap, 557, A124

\bibitem[Brun et al.(2004)]{bru04}
Brun, A., S., Miesch, M. S., Toomre, J., 
2004, 614, 1073 

\bibitem[Burrows et al.(2003)]{bur03}
Burrows, A., Sudarsky, D. \& Hubbard, W. B. 2003, \apj, 594, 545

\bibitem[Charbonneau et al.(2002)]{cha02}
Charbonneau, D., Brown, T. M., Noyes, R. W., \& Gilliland, R. L. 2002, \apj, 568, 377

\bibitem[Choudhuri et al.(1995)]{cho95}
Choudhuri, A. R., Schussler, M., \& Dikpati, M. 1995, \aap, 303, L29

\bibitem[Cooper \& Showman (2005)]{cs05}
Cooper, C. S. \& Showman, A. P. 2005, \apjl, 629, L45

\bibitem[Cranmer \& Saar (2011)]{cs11}
Cranmer, S. R. \& Saar, S. H. 2011, \apj, 741, 54

\bibitem[Dobbs-Dixson \& Lin (2008)]{dl08}
Dobbs-Dixon, I. \& Lin, D. N. C. 2008, \apj, 673, 513

\bibitem[Ehrenreich et al.(2012)]{ehr12}
Ehrenreich, D., Bourrier, V., Bonfils, X., Lecavelier des Etangs, A., H\'ebrard, G., Sing, D. K., Wheatley, P. J., Vidal-Madjar, A., Delfosse, X., Udry, S., Forveille, T. \& Moutou, C. 2012, \aap, 547, A18

\bibitem[Ekenb{\"a}ck et al.(2010)]{eke10}
Ekenb{\"a}ck, A., Holmstr{\"o}m, M., Wurz, P., Grie{\ss}meier, J. M., Lammer, H., Selsis, F. \& Penz, T. 2010, \apj, 709, 670

\bibitem[Fortney et al.(2007)]{for07}
Fortney, J. J., Marley, M. S. \& Barnes, J. M. 2007, \apj, 659, 1661

\bibitem[Garc{\'{\i}}a Mu{\~n}oz (2007)]{gar07}
Garc{\'{\i}}a Mu{\~n}oz, A. 2007, \planss, 55, 1426

\bibitem[Goldstein (1978)]{gol78}
Goldstein, M. L. 1978, \apj, 219, 700 

\bibitem[Heyvaerts \& Priest(1983)]{hp83}
Heyvaerts, J. \& Priest, E. R. 1983, \aap, 117, 220

\bibitem[Holmstr{\"o}m et al.(2008)]{hol08}
Holmstr{\"o}m, M., Ekenb{\"a}ck, A., Selsis, F., Penz, T., Lammer, H. \& Wurz, P. 2008, \nat, 451, 970

\bibitem[Hotta et al.(2012)]{hot12}
Hotta, H., Rempel, M., Yokoyama, T., Iida, Y., \& Fan, Y., \aap, 539, A30

\bibitem[Huang \& Cumming (2012)]{hc12}
Huang, X. \& Cumming, A. 2012, \apj, 757, 47

\bibitem[Ito et al.(2010)]{ito10}
Ito, H., Tsuneta, S., Shiota, D., Tokumaru, M., Fujiki, K. 
2010, \apj, 719, 131

\bibitem[Kopp \& Holzer (1976)]{kh76}
Kopp, R. A. \& Holzer, T. E. 1976, \solphys, 49, 43 

\bibitem[Kudoh \& Shibata (1999)]{ks99} 
Kudoh, T., Shibata, K. 1999, \apj, 514, 493

\bibitem[Kurokawa \& Nakamoto (2014)]{kn14}
Kurokawa, H. \& Nakamoto, T. 2014, \apj, 783, 54

\bibitem[Lammer et al.(2003)]{lam03}
Lammer, H., Selsis, F., Ribas, I., Guinan, E. F., Bauer, S. J. \& Weiss, W. W 2003, \apjl, 598, L121

\bibitem[Landini \& Monsignori-Fossi (1990)]{lm90} 
Landini, M., Monsignori-Fossi, B.C.: 1990, A\&AS, 82, 229

\bibitem[Lecavelier des Etangs et al.(2010)]{lec10}
Lecavelier des Etangs, A., Ehrenreich, D., Vidal-Madjar, A., Ballester, G. E., D\'esert, J. M., Ferlet, R. H\'ebrard, G., Sing, D. K., \& Tchakoumegni, K. O. 2010, \aap, 514, A72

\bibitem[Linsky et al.(2010)]{lin10}
Linsky, J. L., Yang, H., France, K., Froning, C. S., Green, J. C., Stocke, J.T. \& Osterman, S. N. 2010, \apj, 717, 1291

\bibitem[Madhusudhan \& Seager (2009)]{mas09}
Madhusudhan, N. \& Seager, S. 2009, \apj, 2009, 707, 24

\bibitem[Madhusudhan \& Seager (2010)]{mas10}
Madhusudhan, N. \& Seager, S. 2010, \apj, 2010, 725, 261

\bibitem[Matsumoto \& Kitai (2010)]{mk11}
Matsumoto, T. \& Kitai, R. 2010, \apjl, 716, L19

\bibitem[Matsumoto \& Suzuki (2012)]{ms12}
Matsumoto, T. \& Suzuki, T. K. 2012, \apj, 749, 8

\bibitem[Matsumoto \& Suzuki (2013)]{ms13}
Matsumoto, T. \& Suzuki, T. K. 2013, submitted to \mnras

\bibitem[Matthaus et al.(1999)]{mat99} 
Matthaeus, W.H., Zank, G.P., Oughton, S., Mullan, D.J., Dmitruk, P. 
1999, \apjl, 523, L93

\bibitem[Menou (2012)]{men12}
Menou, K. 2012, \apj, 745, 138

\bibitem[Muchmore et al.(1987)]{muc87}
Muchmore, D. O., Nuth, III J. A. \& Stencel R. E. 1987, \apjl, 315, L141

\bibitem[Murray-Clay et al.(2009)]{mur09}
Murray-Clay, R. A., Chiang, E. I. \& Murray, N. 2009, \apj, 693, 23

\bibitem[Perna et al.(2010)]{per10}
Perna, R., Menou, K., \& Rauscher, E. 2010, \apj, 724, 313

\bibitem[Poppenhaeger et al.(2013)]{pop13}
Poppenhaeger, K., Schmitt, J. H. M. M. \& Wolk, S. J. 2013, \apj, 773, 62

\bibitem[Rogers \& Showman (2014)]{rs14}
Rogers, T. M. \& Showman, A. P. 2014, \apjl, 782, L4

\bibitem[Sano et al.(1999)]{san99}
Sano, T., Inutsuka, S., \& Miyama, S. M. 1999, ASSL, 240, 383

\bibitem[Shiota et al.(2012)]{shi12}
Shiota, D., Tsuneta, S., Shimojo, M., Sako, N., Orozco Su\'{o}rez, D., 
Ishikawa, R. 2012, \apj, 753, 157 

\bibitem[Showman \& Guillot (2002)]{sg02}
Showman, A. P. \& Guillot, T. 2002, \aap, 385, 166

\bibitem[Sorahana et al.(2014)]{sor14}
Sorahana, S., Suzuki, T. K. \& Yamamura, I. 2014, \mnras, to be published

\bibitem[Sutherland \& Dopita (1993)]{sd93}
Sutherland, R. S. \& Dopita, M. A. 1993, \apjs, 88, 253

\bibitem[Suzuki (2007)]{suz07}
Suzuki, T. K. 2007, \apj, 659. 1592

\bibitem[Suzuki \& Inutsuka (2005)]{si05}
Suzuki, T. K. \&  Inutsuka, S. 2005, \apjl, 632, L49 

\bibitem[Suzuki \& Inutsuka (2006)]{si06}
Suzuki, T. K. \&  Inutsuka, S. 2006, \jgr, 111, A06101

\bibitem[Suzuki et al.(2013)]{suz13}
Suzuki, T. K., Imada, S., Kataoka, R., Kato, Y., Matsumoto, T., Miyahara, H., 
Tsuneta S. 2013, PASJ, in press

\bibitem[Terasawa et al.(1986)]{ter86}
Terasawa, T., Hoshino, M., Sakai, J. I., \& Hada, T. 1986, \jgr, 91, 4171 

\bibitem[Tian et al.(2005)]{tia05}
Tian, F., Toon, O.B., Pavlov, A. A. \& De Sterck, H. 2005, \apj, 621, 1049

\bibitem[Trammell et al.(2011)]{tra11}
Trammell, G. B., Arras, P. \& Li, Z. Y. 2011, \apj, 728, 152

\bibitem[Trammell et al.(2014)]{tra14}
Trammell, G. B., Li, Z. Y. \& Arras, P. 2014, \apj, in press

\bibitem[Tsuji (1967)]{tsu67}
Tsuji, T. 1967, Late-Type Stars, 260

\bibitem[Tsuji (1973)]{tsu73}
Tsuji, T. 1973, \aap, 23, 411

\bibitem[Tsuneta et al.(2008)]{tsu08} 
Tsuneta, S. et al. 2008, \apj, 688, 1374

\bibitem[Vernezza et al.(1981)]{ver81}
Vernazza, J. E., Avrett, E. H. \& Loeser, R. 1981, \apjs, 45, 635

\bibitem[Vidal-Madjar et al.(2003)]{vid03}
Vidal-Madjar, A., Lecavelier des Enangs, A., D\'esert, J. M., Ballester, G. E., Ferlet, R., H\'ebrard, G. \& Mayor, M. 2003, \nat, 422, 143

\bibitem[Vidal-Madjar et al.(2004)]{vid04}
Vidal-Madjar, A., D\'esert, J. M., Lecavelier des Etangs, A., H\'ebrard, G., Ballester, G. E., Ehrenreich, D., Ferlet, R., McConnell, J. C., Mayor, M. \& Parkinson, C. D. 2004, \apjl, 604, L69

\bibitem[Vidal-Madjar \& Lecavelier des Etangs (2004)]{vl04}
Vidal-Madjar, A. \& Lecavelier des Etangs, A 2004, in ASP Conf. Ser. 321, Extrasolar Planets: Today and Tomorrow, ed. J. Beaulieu, A. Lecavelier des Etangs, \& C. Terquem (San Francisco, CA: ASP), 152

\bibitem[Wu \& Lithwick (2013)]{wl13}
Wu, Y. \& Lithwick, Y. 2013, \apj, 763, 13

\bibitem[Yelle (2004)]{yel04}
Yelle, R. V. 2004, Icarus, 170, 167

\bibitem[Yelle (2006)]{yel06}
Yelle, R. V. 2006, Icarus, 183, 508

\end{thebibliography}
\end{document}